\documentclass[lettersize,journal]{IEEEtran}
\usepackage{amsmath,amsfonts}
\usepackage{algorithmic}
\usepackage{algorithm}
\usepackage{array}
\usepackage{xcolor}
\usepackage[caption=false,font=normalsize,labelfont=sf,textfont=sf]{subfig}

\usepackage{textcomp}
\usepackage{stfloats}
\usepackage{url}
\usepackage{verbatim}
\usepackage{graphicx}
\usepackage{cite}
\usepackage{orcidlink}
\usepackage{threeparttable}

\begin{document}

\title{A 33.6-136.2 TOPS/W  Nonlinear Analog Computing-In-Memory Macro  for Multi-bit LSTM Accelerator in 65 nm CMOS}

\author{
    \IEEEauthorblockN{
        Junyi Yang\orcidlink{0000-0002-5867-4943},~\IEEEmembership{Graduate Student Member,~IEEE},
        Xinyu Luo\orcidlink{0009-0007-5229-8733},
        Ye Ke\orcidlink{0009-0002-9809-1192},
        Zheng Wang,~\IEEEmembership{Member,~IEEE},
        Hongyang Shang\orcidlink{0009-0007-6276-1947},
        Shuai Dong\orcidlink{0009-0007-4807-5094},
        Zhengnan Fu\orcidlink{0009-0009-1235-8521}, 
        Xiaofeng Yang\orcidlink{0000-0003-1400-7994},
        Hongjie Liu\orcidlink{0009-0005-0904-5736},\\
        and Arindam Basu\orcidlink{0000-0003-1035-8770},~\IEEEmembership{Senior Member,~IEEE}\\
        \thanks{
            This work was supported by ITF MSRP grant ITS/018/22MS. \textit{(Corresponding authors: Arindam Basu)} 
            Arindam Basu is with Department of Electrical Engineering and the State Key Laboratory of Terahertz and Millimeter Waves, City University of Hong Kong, Hong Kong, China (e-mail: arinbasu@cityu.edu.hk).    
            Junyi Yang, Xinyu Luo, Ye Ke, Hongyang Shang, Shuai Dong, Zhengnan Fu are with Department of Electrical Engineering, City University of Hong Kong, Hong Kong, China. 
            Zheng Wang is with the Shenzhen Institutes of Advanced Technology, Chinese Academy of Sciences, Shenzhen 518055, China.    
            Xiaofeng Yang and Hongjie Liu are with Reexen Technology, Shenzhen 518000, China.
        }
    }
}



\maketitle

\begin{abstract}

The energy efficiency of analog computing-in-memory (ACIM) accelerator for recurrent neural networks, particularly long short-term memory (LSTM) network, is limited by the high proportion of nonlinear (NL) operations typically executed digitally. To address this, we propose an LSTM accelerator incorporating  an ACIM macro with reconfigurable (1-5 bit) nonlinear in-memory (NLIM) analog-to-digital converter (ADC) to compute NL activations directly in the analog domain using: 1)  a dual 9T bitcell with decoupled read/write paths for signed inputs and ternary weight operations; 2) a read-word-line underdrive Cascode (RUDC) technique achieving 2.8× higher read-bitline dynamic range than single-transistor designs (1.4× better over conventional Cascode structure with 7× lower current variation); 3) a dual-supply 6T-SRAM array for efficient multi-bit weight operations and reducing both bitcell count (7.8×) and latency (4×) for 5-bit weight operations. We experimentally demonstrate 5-bit NLIM ADC for approximating NL activations in LSTM cells, achieving average error $<$1 LSB. Simulation confirms the robustness of NLIM ADC against temperature variations thanks to the replica bias strategy. Our design achieves 92.0\% on-chip inference accuracy for a 12-class keyword-spotting task while demonstrating 2.2× higher system-level normalized energy efficiency and 1.6× better normalized area efficiency than state-of-the-art works. The results combine physical measurements of a macro unit—accounting for the majority of LSTM operations (99\% linear and 80\% nonlinear operations)—with simulations of the remaining components, including additional LSTM and fully connected layers.
\end{abstract}


\section{Introduction}
Conventional von Neumann-based deep neural network (DNN) accelerators face critical performance bottlenecks due to extensive data movement between processor and memory \cite{lee202328}, \cite{diao2024multiply}. 
Computing-in-memory (CIM), which integrates processor and memory together, has gained popularity as a solution to address these issues \cite{wu2024integer},\cite{ju202465},\cite{ liu202033},\cite{wu20238b}. It is a better architecture for the massively parallel execution of multiply-accumulate (MAC) operations, enabling computations to be performed directly within the memory unit itself.  This design effectively eliminates the need for data movement between the processor and memory, thereby enhancing efficiency. Among the reported CIM accelerators, analog computing in-memory  (ACIM) utilizing CMOS-based memory devices, such as static random access memory (SRAM), has demonstrated significant potential in enhancing energy and area efficiencies\cite{ diao2024multiply},\cite{ chen202115},\cite{chi202116}. However, SRAM-based ACIM faces challenges due to multi-bit weight, limited signal margin of read bitline voltage ($V_{RBL}$) and analog-to-digital converter (ADC)  overhead\cite{ wu2024integer},\cite{yu202265},\cite{ si2019twin}. These challenges require urgent attention to enhance the performance of ACIM systems.

\textbf{Challenge one:  Multi-bit ternary weight and  signed input.} 
One of the current challenges for ACIM is to support multi-bit weights\cite{wang2023charge},\cite{yue202014}. In many SRAM-based ACIM scheme, multibit weights are typically stored in a group of SRAM cells because each SRAM cell can only store one digital bit. To implement multi-bit weights, the first method is to change the amplitude of the  read word line (RWL) signals using digital-to-analog converters (DACs) \cite{wang2023charge},\cite{yue202014}, which increases hardware overhead and suffers from NL relationship from RWL-voltage to $V_{RBL}$ \cite{wang2023charge}.  The second method is to change the pulse width of RWL \cite{dong2025topkima},\cite{biswas2018conv}. However, this approach results in a decrease in throughput by 2\textsuperscript{$n_w-1$} times (where $n_w$ denotes weight precision). The third method, utilized in \cite{yu202265},\cite{yu2024dual}, avoids a decrease in throughput but requires $2\textsuperscript{$n_w$}-1$ cells for $n_w$-bit weight, leading to substantial hardware overhead. Another issue is the handling of signed inputs. Currently, many CIMs \cite{diao2024multiply},\cite{ju202465},\cite{liu202033},\cite{wu20238b},\cite{wang2023charge},\cite{yue202014} support only unsigned inputs, as they target convolutional neural networks (CNNs) and multi-layer perceptrons (MLPs) with ReLU-generated non-negative values. However, for recurrent neural networks (RNNs) and attention-based neural network, the inputs are signed. Therefore, it is crucial to design CIM systems that can effectively support signed inputs to meet the requirements of these more complex neural network architectures.

\textbf{Challenge two: Limited signal margin and linearity.} Boosting MAC operations via multiple RWLs improves EE and throughput but widens MAC voltage range, reducing signal margin under fixed dynamic range (DR) of $V_{RBL}$ \cite{yu202265}. 
To address this limitation, a source degeneration architecture is employed to enhance the signal margin and increase the voltage swing range \cite{si2019twin}. This approach achieves a signal margin that is 1.44× larger than that of conventional 6T-SRAM cells while maintaining the same precision in MAC operations.

\textbf{Challenge three: ADC for ACIM.}
Designing an efficient ADC for ACIM  presents significant challenges due to the significant overhead in both area and energy consumption \cite{yue202014},\cite{kim2023neuro}. In-memory (IM) ADC is proposed to reduce the overhead of traditional ADC and convert analog MAC signals to digital code simultaneously \cite{yu202265},\cite{dong2025topkima},\cite{yu202016k}. However, IM ADC incurs a substantial area overhead for calibration, since the number of bitcells needed for calibration is equivalent to that of the ADC, effectively doubling the memory requirement \cite{yu202265}.
\begin{figure*}[!t]
  \centering
  \includegraphics[width=1\linewidth]{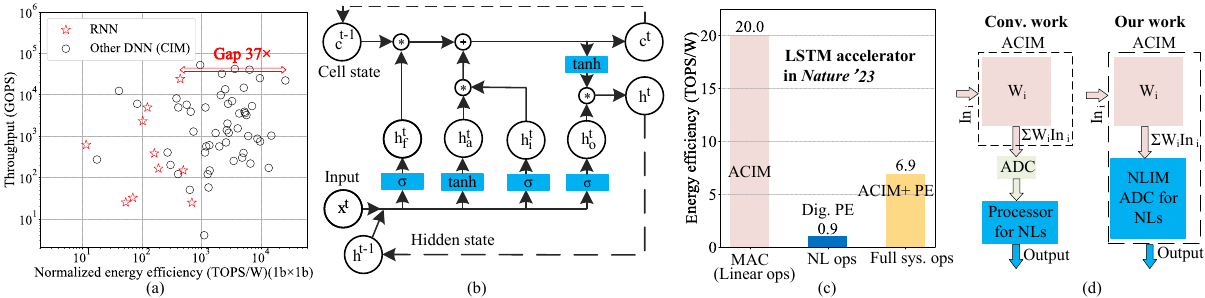}
  \caption{Limitations and solutions of current CIM for RNNs: (a)  A survey of RNN accelerators and other CIM-based DNN accelerators (All energy efficiencies are normalized to 1-bit input and 1-bit weight, according to this formula \cite{song20234}: Normalized EE=EE × input precision × weight precision). (b) One LSTM cell with a large number of nonlinear activations.(c) Energy efficiency of various sub-parts of the LSTM accelerator in previous work (\textit{Nature'23},\cite{ambrogio2023analog}).  (d) Architecture comparison of our proposed method with the conventional method for LSTM accelerator.}
  \label{Introduction_fig}
\end{figure*}

Most importantly, this improvement of efficiencies by CIM does not extend to the full system when implementing RNNs, such as long short-term memory (LSTM) networks. Fig. \ref{Introduction_fig}(a) illustrates normalized energy efficiency (EE) based on input and weight resolutions and throughput of current RNN accelerators \cite{conti2018chipmunk,yin20171,shin201714,tambe202216,ambrogio2023analog,le202364,dbouk20200,guo20216,kadetotad20208,guo20195} in comparison to those of recent CNN and MLP accelerators using CIM. With the CNN accelerator achieving a normalized EE of 25840 TOPS/W \cite{yan202428}  compared to just normalized 696.5 TOPS/W \cite{kadetotad20208} for the RNN accelerator, this 37× gap highlights the limitation of current CIM in improving the EE of RNN accelerators. The limitation in EE improvements for RNNs, particularly LSTM, stem from the high proportion of nonlinear (NL) operations as shown in Fig. \ref{Introduction_fig}(b) typically executed digitally. These NL operations, such as the sigmoid and hyperbolic tangent functions used as activation functions, are computationally intensive and can become a significant bottleneck in the overall processing efficiency. For example, \cite{ambrogio2023analog} conducted a comparative analysis about the EEs of various sub-parts of an ACIM-based LSTM accelerator as depicted in Fig. \ref{Introduction_fig}(c). In this system, the linear operations (MAC) are executed by ACIM with high EE (20 TOPS/W), while the NL operations are performed by the digital processor resulting in a low EE (0.9 TOPS/W). The overall EE of the system (6.9 TOPS/W) is diminished by a factor of 2.9 in comparison to 20 TOPS/W due to the significantly lower EE associated with the NL operations in the LSTM cell. Addressing the computational bottleneck induced by NL operations is critical, especially since compact LSTM architectures demonstrate superior parameter efficiency over Transformer-based networks while maintaining competitive performance in sequential modeling tasks such as speech recognition \cite{kadetotad20208} and char prediction \cite{le202364}.

There are several methods to calculate NL operations in LSTM cell, such as piecewise linear approximations\cite{kadetotad20208},\cite{chong2021efficient},\cite{feng2021high}, cordic \cite{raut2020cordic}, lookup table \cite{xie2020twofold} , quadratic polynomial approximation \cite{li2020low}. However, these methods all compute each NL operation serially in the digital domain, which increases system latency and reduces energy efficiency.  

Recent studies\cite{yang2025efficient, xie2025rac} have attempted to compute NL activation functions in the analog domain. A resistive random-access memory (RRAM)-based NL ADC was proposed in \cite{yang2025efficient} to approximate the NL activation function; however, this work only validated the accuracy of the NL functions in the presence of RRAM programming errors, without integrating it directly with a MAC unit in silicon. This is primarily due to the difficulty of integrating RRAM arrays with complex CMOS peripheral circuits. Additionally, two major drawbacks are noted: the significant programmed error and stuck-at-fault ratio of RRAM devices considerably degrade network accuracy (88.5\% for 12-class under 5-bit ADC, Google Speech Commands Dataset (GSCD)), and the charge-domain in-memory computing architecture adopted employs integrators that account for more than half of the total system power, thereby reducing overall energy efficiency. In \cite{xie2025rac}, analog circuitry utilizing Taylor approximation to fit nonlinear activation functions was introduced. While this approach has been validated at the module level, it has not been integrated with CIM array to demonstrate overall performance in silicon. Moreover, the module occupies a considerable area of 1218 $\mu m^2$, making it impractical to integrate one module per CIM column to achieve parallel computation. Another limitation is that its output remains in the analog domain, which the authors claim can help to remove AD/DA overheads and be passed directly to the next layer. However, this is only true for small networks where direct connections can be made–in most practical cases, the output has to be stored in local memory for further digital processing (such as element wise products in LSTM) or data staging (creating feature maps in CNN for use by the next layer)\cite{ambrogio2023analog}. This necessitates an additional ADC to quantize the output signal for communication with the external digital controller. This extra ADC not only increases system overhead but also further degrades the accuracy of the NL approximation due to quantization errors. Further, this method \cite{xie2025rac} uses sub-threshold techniques for log-conversion–when used in an array, such approaches are known to suffer from exacerbated mismatch due to the exponential dependence of the output current on threshold voltage mismatch. Some other works \cite{yin2020xnor,kneip2023impact} have designed ADCs with programmable range to better match with the distribution of MAC values; however, they cannot be programmed to perform arbitrarily nonlinear operations. Consequently, it is imperative to explore alternative approaches that can perform NL operations more efficiently to enhance the overall efficiency of LSTM accelerators. 

To address the aforementioned issues, the main contributions of this work are summarized as follows:

1) A dual 9T bitcell is proposed to support signed inputs and ternary weights, combined with a dual-supply 6T-SRAM for efficient implementation for multi-bit weights, which reduces bitcell count by 7.8× and latency by 4× compared to prior methods \cite{yu202265},\cite{dong2025topkima} for 5-bit weight operations.

2) We propose a read-word-line under drive Cascode (RUDC) technique that increases DR of $V_{RBL}$ by 2.8× versus prior work \cite{yu202265}  and  1.4× versus conventional Cascode structure with 7× improvement of discharge current linearity. 

3) We propose an innovative SRAM-based reconfigurable nonlinear in-memory (NLIM) ADC that directly computes LSTM activation functions (sigmoid/tanh) in the analog domain as illustrated in Fig. \ref{Introduction_fig}(d). The hardware-implemented NLIM ADC exhibits an error of less than 1 LSB when approximating the sigmoid and tanh functions, overcoming digital computation bottlenecks while maintaining temperature robustness due to replica bias. Our NLIM ADC (including the NL controller) achieves a 3.3× improvement in normalized area over the SAR ADC in \cite{zhang2023macc} with an NL processor. It also achieves a 3.7× improvement compared to the linear in-memory ADC in \cite{yu202265} with an NL processor.

4) Our keyword-spotting (KWS) system achieves 92\% (better than previous SOTA: 91.8\% \cite{tan202417}) on-chip inference accuracy (12-class, GSCD) , enabled by quantization and hardware nonidealities aware training, with 2.2× higher system-level normalized EE and 1.6× better  normalized area efficiency than SOTA RNN accelerators\cite{guo20216},\cite{guo20195}.

\section{Macro structure}
\label{sec:Macro structure and software}
To solve the above challenges, 1) we propose dual 9T SRAM bitcell to implement signed input and dual-supply 6T-SRAM array to implement multi-bit weight efficiently. 2) We utilize RUDC to enhance DR of $V_{RBL}$ and linearity of discharge current of bitcell.  3) We propose NLIM ADC to reduce the huge hardware overhead of ADC calibration bitcells and enables the parallel computation of NL operations in the analog domain.

\begin{figure*}[!t]
  \centering
  \includegraphics[width=1.0\linewidth]{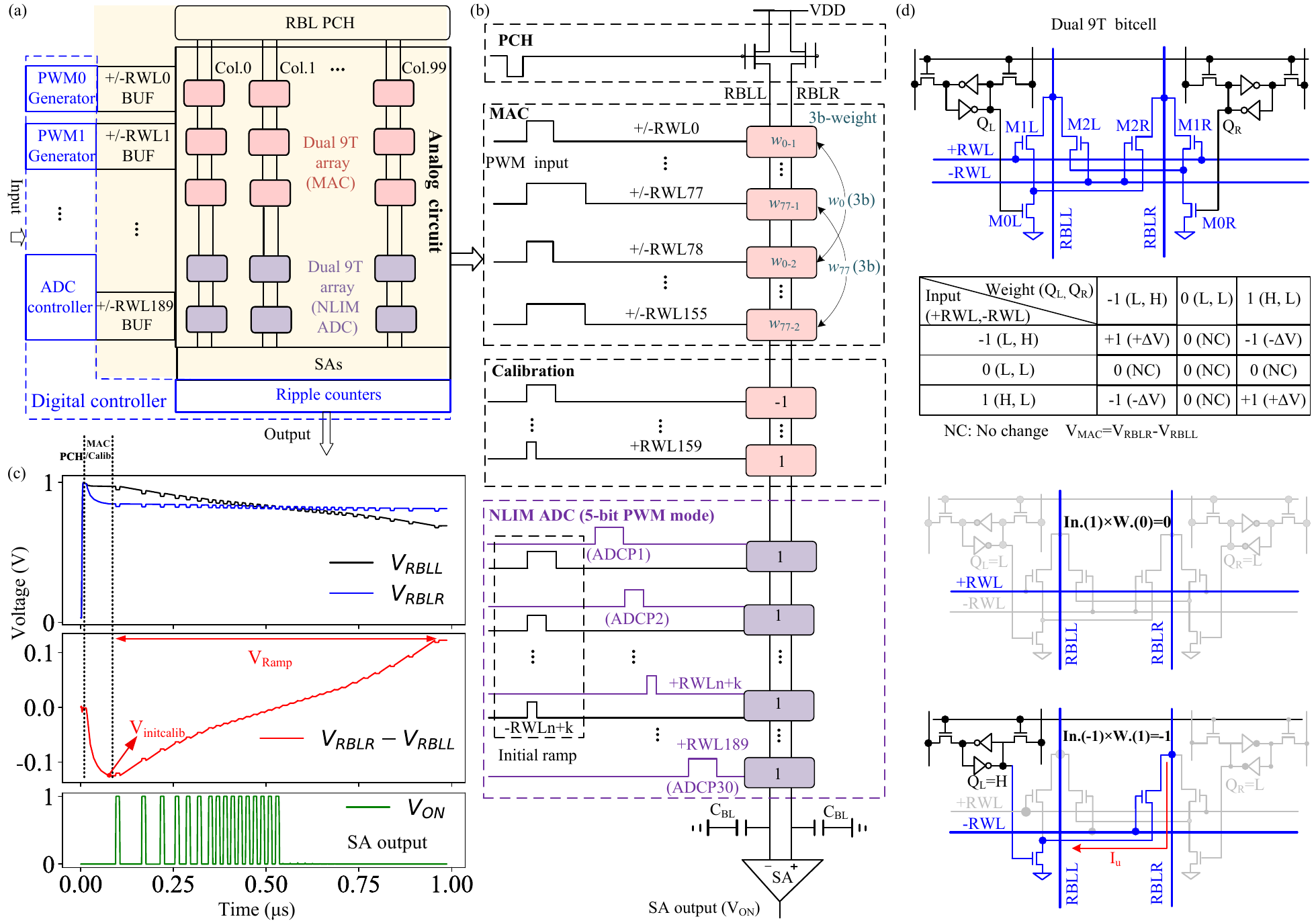}
  \caption{ Hardware block diagram: (a) Macro architecture in the test chip (BUF: buffer). (b) Circuit timing diagram of one column (PCH: precharge). (c) Voltage waveforms diagrams of RBL and outputs of SA based on post-layout simulation. (d) In-memory ternary multiplication of proposed dual 9T SRAM bitcell.}
  \label{chip_overall_architecture}
\end{figure*}

\subsection{Proposed System Architecture and Application}
\label{sec:Macro Structure and Overview}

 The top-level schematic of the CIM and NLIM ADC based on SRAM is depicted in Fig. \ref{chip_overall_architecture}(a), which includes $160\times 100$ dual 9T SRAM bitcells for MAC and calibration, $30\times 100$ bitcells for references of the reconfigurable NLIM ADC, sense amplifiers(SAs), RWL buffers, pulse-width modulation (PWM) generators, NLIM ADC controller, 
 and ripple counters. To clearly demonstrate how it works, the timing of MAC operations using PWM mode for multi-bit input and a set of NLIM ADC timing diagram are shown in Fig. \ref{chip_overall_architecture}(b). PWM inputs, calibration inputs, and intial ramp inputs of NLIM ADC part are applied simultaneously (Fig. \ref{chip_overall_architecture}(b)) to create $V_{MAC}=V_{RBLR}-V_{RBLL}$ on the bitlines (using current-mode operation). Fig. \ref{chip_overall_architecture}(c) shows voltage waveforms of two $V_{RBL}s$ and their difference based on post-layout simulation. After MAC, calibration, and initial ramp, $V_{MAC}$ is in the position of $V_{initcalib}$ in Fig. \ref{chip_overall_architecture}(c). Next, the NLIM ADC part creates a reference decreasing voltage on RBLL every clock cycle using ADC pulses (ADCP1-ADCP30 in Fig. \ref{chip_overall_architecture}(b)), which effectively increases the ramp voltage ($V_{Ramp}$) (Fig. \ref{chip_overall_architecture}(c)). The NLIM ADC controller is used to generate the RWLs pulse of the NLIM ADC part shared by all columns of NLIM ADC. Therefore, each of the $100$ SAs can simultaneously compare $V_{RBLL}$ with $V_{RBLR}$. The ripple counters convert the output thermometer code of the SAs into  binary codes. The system clock frequency in our implementation is 100 MHz.
\subsection{Dual 9T SRAM-based In-Memory Ternary Multiplication and Accumulation}
\label{sec:Dual 9T SRAM-based In-Memory Binary Multiplication and Accumulation}
 A dual 9T SRAM bitcell has been employed for the implementation of the macro as illustrated  in Fig. \ref{chip_overall_architecture}(d) (top).  This bitcell facilitates ternary multiplication of a signed input with a ternary weight using a decoupled read path, which includes the six blue NMOS transistors (three transistors on each side as depicted  in Fig. \ref{chip_overall_architecture}(d)).
 
For positive inputs, the high voltage is applied to the +RWL, while for negative inputs, the high voltage is applied to the -RWL. Ternary weights (-1($Q_{L}$=L, $Q_{R}$=H); 0(($Q_{L}$=L, $Q_{R}$=L)) and +1(($Q_{L}$=H, $Q_{R}$=L))) can be stored in the dual 6T-SRAM bitcells. Fig. \ref{chip_overall_architecture}(d) (bottom) depicts two possible multiplication cases with different input and weight values. When the weight is 0 ($Q_{L}$=L, $Q_{R}$=L), no current discharges the RBLL or RBLR, regardless of the inputs. Conversely, when the weight is -1/1, either RBLL or RBLR is discharged via the corresponding NMOS transistor, where the gate is linked to +RWL/-RWL with a high voltage. The ternary multiplication result is encoded in the voltage difference ($V_{RBLR}$-$V_{RBLL}$) as shown in the table in Fig. \ref{chip_overall_architecture}(d). The layout of the dual 9T bitcell occupies 3.6 $\mu m$$\times$1.9 $\mu m$. To ensure a fair comparison, we  normalize the bitcell size  in Tab. \ref{tab:bitcell_comparison}, taking into account the precision of inputs and weights as well as the process node.  After size normalization, our bitcell demonstrates comparative advantages in normalized size, data type support, and DR of $V_{RBL}$ as shown in Tab. \ref{tab:bitcell_comparison}. 
 
\begin{table}[t!]
  \centering
  \caption{Comparison  of different SRAM bitcells for ACIM \\ (TCASI'25\cite{dong2025topkima}, JSSC'22\cite{yu202265}, JSSC'23\cite{zhang2023macc}) \label{tab:bitcell_comparison}}
  \includegraphics[width=\columnwidth]{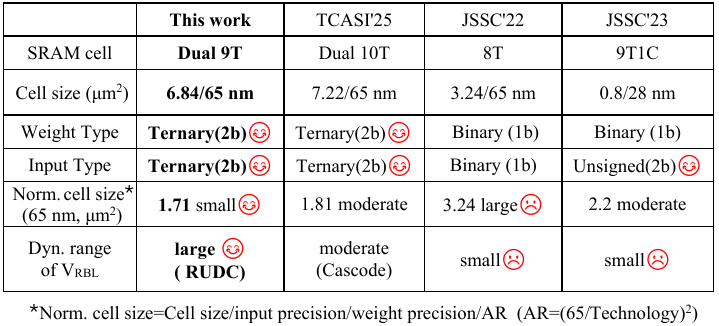} 
\end{table}

 Accumulation occurs due to all bitcells in a column simultaneously discharging the bitline capacitor $C_{BL}$. The MAC value is directly proportional to the voltage drop induced by the capacitor discharge process as described in the formula: $\begin{aligned}V_{M A C} & =V_{R B L R}-V_{R B L L} =\frac{T I_u}{C_{BL}} \sum_{i=0}^{n_{row}-1} \operatorname{In}_i W_i\end{aligned}$, where  $T$ and $I_u$ are the clock period and the unit discharge current of a single bitcell.

\subsection{Dual-supply for Implementation of Multi-bit Weight}
\label{sec:Implementation of Multi-bit Weight}
\begin{figure*}[!t]
  \centering
  \includegraphics[width=1.0\linewidth]{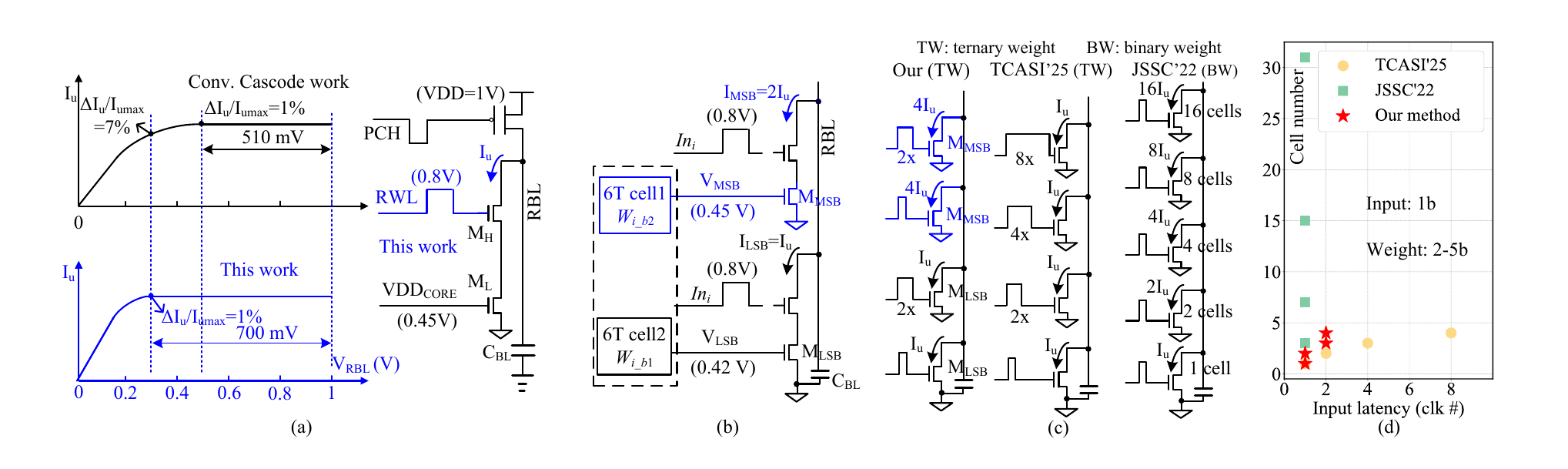}
  \caption{ (a) RUDC for enhanced linearity and DR. (b) Implementation of multi-bit weight using dual-supply for 6T-SRAM array. (c) Three methods (proposed, TCASI'25 \cite{dong2025topkima}, JSSC'22 \cite{yu202265}) for implementing 5-bit signed weight. (d) Comparison of input latency and cell number for implementing multi-bit weight.}
  \label{Implementation of Multi-bit Weight}
\end{figure*}
We employ dual-supply 6T-SRAM array denoted as ( $V_{\text{MSB}}$ ) for the Most Significant Bit (MSB, $W_{i_{-} b 2}$) and  $V_{\text{LSB}}$ for the Least Significant Bit (LSB, $W_{i_{-} b 1}$) to facilitate the mapping of the MSB and LSB of weights to two discharge current $I_{MSB}$ and $I_{LSB}$ ($I_{MSB}$ = $n_{BWR}I_{LSB}$), where $n_{BWR}$ denotes the bit weighting ratio (BWR) and is selected according to the desired weight resolution. Fig. \ref{Implementation of Multi-bit Weight}(b) shows a case for 3-bit weights and $n_{BWR}=2$ where the weight is represented as 2 bits, excluding the sign bit ( $W_i: W_{i_{-} b 2} W_{i_{-} b 1} $). The MAC and calibration arrays of the 6T-SRAM bitcells consists of 160 rows, which are evenly divided into two arrays: the upper 80 rows are powered by  $V_{\text{MSB}}$  for MSB weights, while the lower 80 rows are powered by   $V_{\text{LSB}}$  for LSB weights.
\begin{figure}[htbp]
  \centering
  \includegraphics[width=1\linewidth]{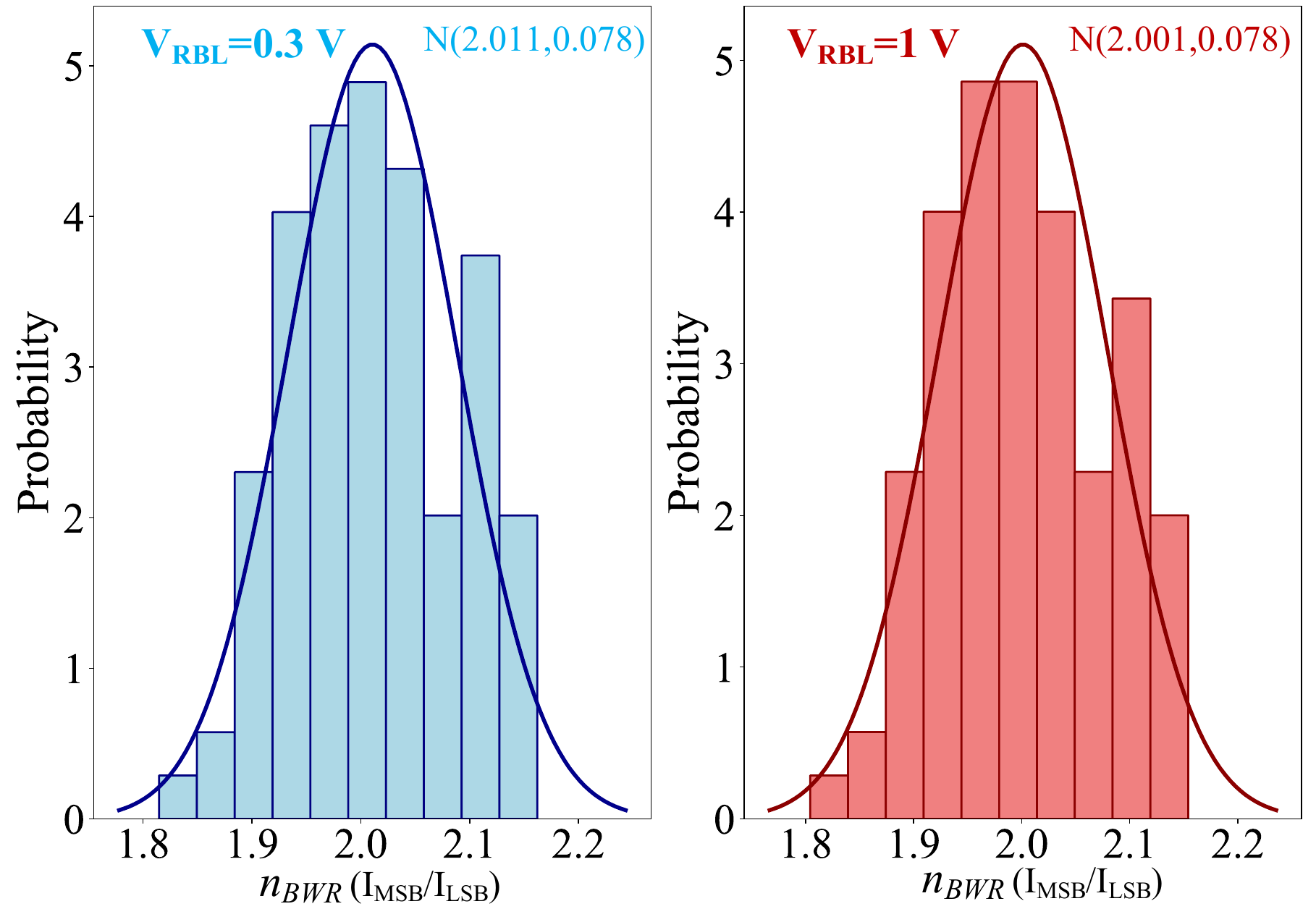}
  \caption{Monte Carlo simulations for the ratio $n_{BWR}=2$ under the conditions of $V_{MSB}$ = 0.45 V and $V_{LSB}$ = 0.42 V. }
  \label{iu_ratio}
\end{figure}

\begin{table}[]

  \caption{Configuration settings for 3-bit and 5-bit weight in the proposed approach (1-bit input)}

\begin{tabular}{|l|l|l|}
\hline
                 & 3-bit weight & 5-bit weight \\ \hline
Bitcell number       & 2            & 4            \\ \hline
Latency (clk number) & 1            & 2            \\ \hline
$n_{BWR}$ $(I_{MSB}/I_{LSB})$        & 2            & 4            \\ \hline
$\Delta VDD_{CORE}$ ($V_{MSB}-V_{LSB}$) (mV)   & 30           & 60           \\ \hline
\end{tabular}
  \label{tab:3-5bit weight}

\end{table}

\begin{table}[]
    \centering
 \caption{Weight value stored per bitcell (for 5-bit weights)  in the proposed approach}
\begin{tabular}{|l|llll|}
\hline
{Signed 5-bit   weight} & \multicolumn{4}{l|}{ Weight values of bitcells in   different rows}                                          \\ \cline{2-5} 
                                       & \multicolumn{1}{l|}{Bitcell0} & \multicolumn{1}{l|}{Bitcell1} & \multicolumn{1}{l|}{Bitcell2} & Bitcell3 \\ \hline
11111 (-15)                             & \multicolumn{1}{l|}{-1}   & \multicolumn{1}{l|}{-1}   & \multicolumn{1}{l|}{-1}   & -1   \\ \hline
11110 (-14)                             & \multicolumn{1}{l|}{-1}   & \multicolumn{1}{l|}{-1}   & \multicolumn{1}{l|}{-1}   & 0    \\ \hline
00000 (0)                               & \multicolumn{1}{l|}{0}    & \multicolumn{1}{l|}{0}    & \multicolumn{1}{l|}{0}    & 0    \\ \hline
01111 (+15)                             & \multicolumn{1}{l|}{1}    & \multicolumn{1}{l|}{1}    & \multicolumn{1}{l|}{1}    & 1    \\ \hline
\end{tabular}
  \label{tab:5bit weight implementing}
\end{table}

For subthreshold MOSFETs ($M_{MSB}, M_{LSB}$ in Fig. \ref{Implementation of Multi-bit Weight}(b)), the 2:1 ratio ($I_{MSB}/I_{LSB}$), i.e. $n_{BWR}=2$ requires $\Delta VDD_{CORE}=V_{MSB}-V_{LSB}=\frac{U_T}{\kappa} \ln 2 \approx$ 26 mV, where ${U_T}$ is thermal voltage (26 mV @300 K, $\kappa \approx0.7$ is the inverse of sub-threshold slope factor). Utilizing this voltage difference and the fixed ($V_{\text{LSB}}$=0.42 V), we can readily determine the appropriate $V_{\text{MSB}}$. To estimate the effect of mismatch, we performed two Monte Carlo simulations for the ratio $n_{BWR}=2$ under the conditions of $V_{MSB}$ = 0.45 V and $V_{LSB}$ = 0.42 V with $V_{RBL}$ = 0.3 V, 1 V respectively. The results in Fig. \ref{iu_ratio} show that the mean and standard deviation of $n_{BWR}$ change very little as $V_{RBL}$ varies. The $n_{BWR}$ follows a normal distribution N(2.011, 0.078) ($V_{RBL}$ = 0.3 V) with the variation seen to degrade classification accuracy by only $0.21\%$. We also experimentally verified that these two voltage values ($V_{\text{MSB}}$=0.45 V, $V_{\text{LSB}}$=0.42 V) enable the MAC values of the MSB and LSB arrays to be approximately doubled when the weights and input values of both arrays are identical. More accurate ratios can be obtained by calibration as described in Section \ref{sec:weight_calib}. 

Fig. \ref{Implementation of Multi-bit Weight}(c) shows three methods for implementing 5-bit weight ($n_{BWR}=4$ in this case) using PWM input and multi-cell. While PWM methods incur penalties in latency, multi-cell methods incur penalties in area. Tab. \ref{tab:3-5bit weight} shows the different settings of MSB and LSB bitcell power supplies for these two cases (3-bit/5-bit weight). To clearly illustrate how our four bitcells achieve the representation of a 5-bit weight, several examples of signed 5-bit weights are presented in Tab. \ref{tab:5bit weight implementing}, which shows the distinct ternary weight values stored in the four bitcells according to different numerical values of the 5-bit weight.
 Fig. \ref{Implementation of Multi-bit Weight}(d) compares the number of bitcells and the latency required to achieve 2 to 5 bits of weight with those reported in the previous two articles \cite{yu202265},\cite{dong2025topkima}. Dual-supply for 6T-SRAM array are leveraged for multi-bit weights, achieving reductions in cell number/area by 7.8×/3.7×  compared to\cite{yu202265} and latency by 4× compared to \cite{dong2025topkima}, respectively, when implementing 5-bit weights.

Another method for implementing multi-bit weights using binary capacitors was proposed in references \cite{kneip2023impact},\cite{sinangil20207}. However, this approach presents issues of large area occupied by capacitors (e.g. \cite{kneip2023impact} uses $\approx17\%$ of bitcell array), limited resolution and limited configurability. In comparison to these methods, despite requiring an additional power supply, our design offers key advantages over these methods by eliminating the need for large capacitors and enabling flexible configuration of the supported weight bit-width. Since the power drawn by this supply is less than 3 $\mu$W, we estimate a simple reference generator with buffer will suffice to provide this adding negligible overheads in power and area. Note that our dual-supply approach is applicable to other bitcell designs as well. If signed inputs are not necessary, we can use foundry supplied 8T cells with smaller footprint.

\subsection{RWL Underdrive enabled Cascode for enhanced DR and Linearity}
\label{sec:RWL Underdrive enabled Cascode}
Word line underdrive is typically used in SRAM for improving static noise margin. In this work, we achieve high linearity of $I_u$ and large DR of $V_{RBL}$ by RUDC technique--the supply voltages for RWL is set to $V_{RWL}=0.8$ V, while $V_{RBL}$ pre-charging voltage is set at $VDD=1$ V and gate voltage of $M_L$ is at $VDD_{core}=0.45$ V as shown in Fig. \ref{Implementation of Multi-bit Weight}(a). The minimum RBL voltage for which $M_H$ acts as a Cascode is given by $V_{min}=V_{RWL}-V_{T1}$ where $V_{T1}$ denotes the threshold voltage of $M_H$. Lower $V_{RWL}$ leads to higher $V_{RBL}$ swing and the lowest RWL voltage is limited by the requirement to keep the lower NMOS $M_L$ in saturation. We obtain DR of $700$ mV on the RBL for a $1\%$ variation in $I_u$ (Fig. \ref{Implementation of Multi-bit Weight}(a))--7× reduction  than the variation observed in $I_u$ over the same range for the conventional case. For a 1\% current variation, the conventional Cascode method achieves a DR of only $\approx$510 mV, while our method reaches $\approx$700 mV, necessitating a DR reduction by a factor of 1.4× (700/510) for the conventional approach to limit current variations to the same level.
Single transistor (in 8T bitcell) based discharge paths have an even smaller DR of $\approx 250$ mV  \cite{yu202265}. Our DR ($700$ mV) and voltage signal margin ($DR/column\; \#$, 3.68 mV) is 2.8×/1.9× higher than the values of single transistor architecture (8T bitcell in \cite{yu202265}, DR = 250 mv, voltage signal margin =1.95 mV).

\section{Nonlinear In-memory ADC }
\label{sec:In-memory Nonlinear ADC}
\begin{figure*}[htbp]
  \centering
  \includegraphics[width=1\linewidth]{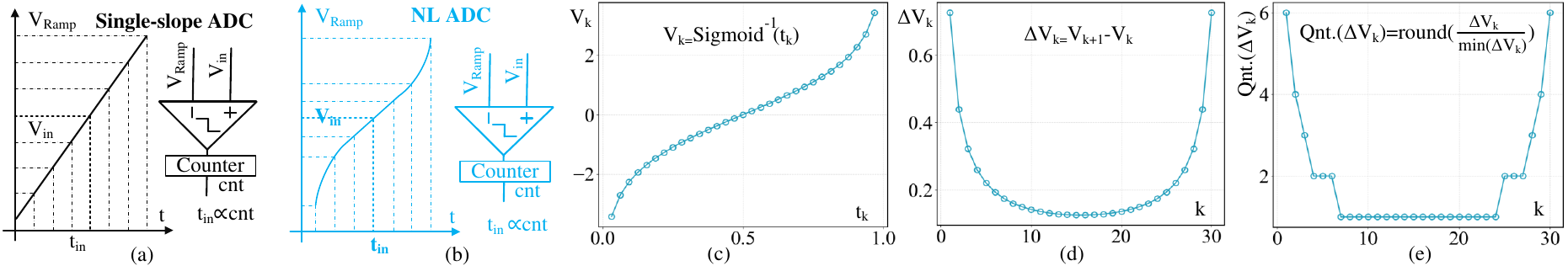}
  \caption{(a) Traditional single-slope ADC. (b) Our Ramp NL ADC. (c) The inverse of the sigmoid function. (d) The value of each step of the ramp voltage $V_k$ denoted by $\Delta V_k$. (e) Integer quantized $\Delta V_k$ for implementation in hardware.}
  \label{Vk and DeltaVk}
\end{figure*}
\subsection{Implementation of NLIM ADC in Hardware}
\label{sec:Implementation of NLIMA}
A traditional single-slope ADC is shown in Fig. \ref{Vk and DeltaVk}(a). Its output is produced by a comparator whose negative input is connected to a time-varying ramp signal, $V_{Ramp}(t)=\frac{1}{k}t$, and the positive input is connected to $V_{in}$, where $f(t)=kt$ is linear. The threshold crossing time of comparator $t_{in}$ ($t_{in}=kV_{in}=f(V_{in})$) can be obtained using a counter in Fig. \ref{Vk and DeltaVk}(a). $t_{in}$ is the output of activations, which is proportional to the counter value. Compare above two equations, we can conclude $V_{Ramp}(t)=f^{-1}(t)$.

For Ramp NL ADC depicted in Fig. \ref{Vk and DeltaVk}(b), the principle is exactly the same as that of the linear one with the only difference that $V_{Ramp}(t)=f^{-1}(t)$ is a NL curve rather than a line. $t_{in}$ still equals $f(V_{in})$ when $V_{Ramp}(t)=V_{in}$, where $f()$ is NL activation, such as sigmoid, tanh. Take the sigmoid function ($f(t)=1/(1+e^{-t})$) and n-bit NLIM ADC as an example. To extract the step of the generated ramp voltages, we first obtain the inverse of the sigmoid function ($f^{-1}(t)=\ln{\frac{t}{1-t}}$) as depicted in Fig. \ref{Vk and DeltaVk}(c). This function is exactly the $V_{Ramp}$ that needs to be generated during the conversion process. The methodology for generating the NL curve of $V_{Ramp}$ using NLIM ADC is presented in the following text (5 steps).

\textbf{1)} $2^n -1$ equidistant sample points ($t_{k}$,$V_{k}$) are extracted from the inverse sigmoid characteristic curve shown in Fig. \ref{Vk and DeltaVk}(c) (here n=5). 

\textbf{2)} The voltage differences $\Delta V_{k}=V_{k+1}-V_{k}$ are computed and plotted in Fig. \ref{Vk and DeltaVk}(d).


\textbf{3)} Each $\Delta V_{k}$ is quantized using the relation $Qnt.(\Delta V_{k})=round(\frac{\Delta V_{k}}{min(\Delta V_{k})})$, where $min(\Delta V_{k})$ represents the minimum voltage difference. The resulting quantized values are shown in Fig. \ref{Vk and DeltaVk}(e).

\begin{figure}[!t]
  \centering
  \includegraphics[width=1\linewidth]{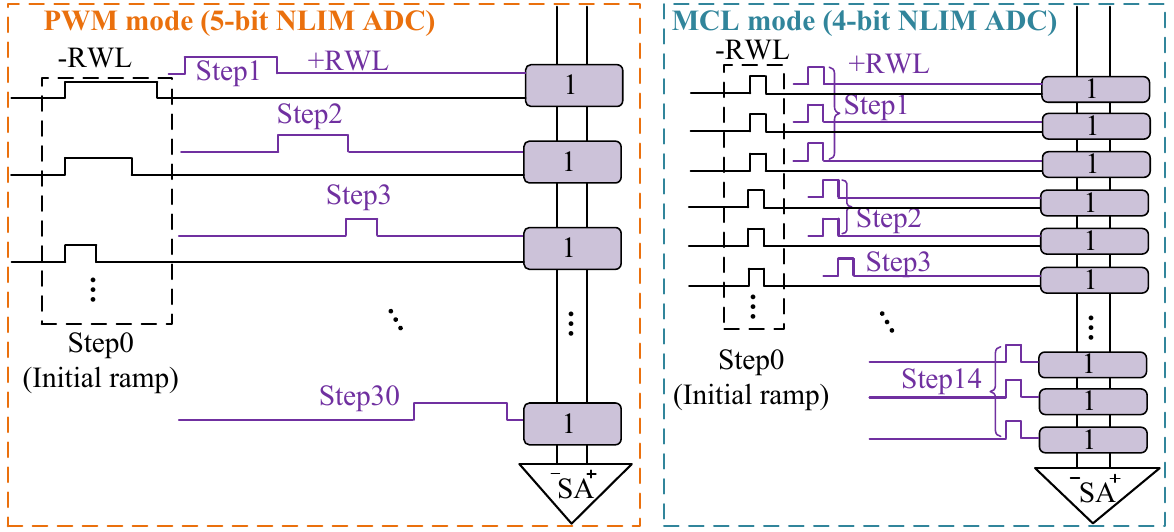}
  \caption{PWM mode and multi-cell mode for implementation of NLIM ADC.}
  \label{two_modes}
\end{figure}

\textbf{4)} Our NLIM ADC generates ramp voltage $V_{Ramp}$ with step size: $\Delta V^k_{Ramp}=\frac{I_uT}{C_{BL}}Qnt.(\Delta V_{k})$. Two operating modes are implemented (Fig. \ref{two_modes}): 1) PWM Mode: single bitcell per step, where pulse width $T_{PWM}=Qnt.(\Delta V_{k})T$ controls $\Delta V^k_{Ramp}$; 2) Multi-Cell (MCL) Mode:  multi-bitcells ($n_{cell}=Qnt.(\Delta V_{k})$) activated per step controlling $\Delta V^k_{Ramp}$, each step for one clock cycle. Tab. \ref{tab:different modes and resolutions} presents a comparison of the two NLIM ADC operating modes (5-bit and 4-bit resolutions) in terms of latency and bitcell number.

\setcounter{equation}{3}
\textbf{5)} The ramp voltage generated at the $p$-th clock cycle ($V^p_{Ramp}$) is given by the following equation. 

$V_{Ramp}^p=V_{initcalib}+\sum_{k=1}^p\Delta V^{k}_{Ramp} \\ =-\frac{I_uT}{2C_{BL}}\sum_{k=1}^{2^n-2} Qnt.(\Delta V_{k})+\frac{I_uT}{C_{BL}}\sum_{k=1}^p Qnt.(\Delta V_{k}) \quad (3)$ where $V_{initcalib}$ is the initial voltage of  $V_{Ramp}$ as shown in Fig. \ref{chip_overall_architecture}(c), which  can be efficiently generated using only $2^{n-1}-1$ bitcells (15 for the 5-bit case depicted) with identical pulse width as the first 15  bitcells of NLIM ADC as depicted in Fig. \ref{two_modes} (PWM mode). In this configuration, all 15 -RWL signals are activated simultaneously, causing RBLR (in Fig. \ref{chip_overall_architecture}(c)) to discharge and generate the $V_{initcalib}$. Then, +RWL signals are enabled sequentially in a step-by-step fashion, allowing the RBLL to discharge producing an increasing voltage difference between the two RBLs named $V_{Ramp}$ in Fig. \ref{chip_overall_architecture}(c).

As derived from Equation (3), our architecture can generate NL characteristic curves of varying shapes by programming different voltage values, approximating various activations. The NLIM ADC design maintains full flexibility to operate in linear mode (where all $Qnt.(\Delta V_{k})$ values are identical) when needed. The implementation utilizes 30 rows of bitcells, supporting ADC resolutions configurable from 1 to 5 bits.

Tab. \ref{tab:ADC_comparison} presents a comprehensive comparison between our NLIM ADC and state-of-the-art ADC implementations  \cite{yu202265},  \cite{zhang2023macc}, \cite{lee202328}. To ensure a fair comparison of area efficiency across different MAC array sizes, ADC resolutions, and technology nodes, we normalized the area of all ADCs. The total area of each ADC design is first divided by the number of columns in the MAC array to represent the area per column. This per-column ADC area is then normalized to an equivalent 5-bit resolution in a 65 nm technology node using the scaling formula provided in Tab. \ref{tab:ADC_comparison}.

For a comprehensive comparison, we also include the area of the NL control logic. In our design, the total area of the NL controller is divided by the number of columns, as a single controller is shared across all columns. For other linear ADCs, we add the area of one NL processor (783.4 $\mu m^2$ in 65 nm based on \cite{lyu2021ml}).

Finally, we compared the combined normalized area of one ADC plus the corresponding NL unit. As shown in Tab. \ref{tab:ADC_comparison}, the total normalized one ADC area plus NL area of our design is 3.3× better than that of the most area-efficient prior work  \cite{zhang2023macc} and 3.7× better than that of the  linear in-memory ADC in \cite{yu202265}.


\begin{table}[t!]
  \centering
  \caption{Comparison of different modes and resolutions. }
\begin{tabular}{|l|l|l|l|l|}
\hline
\begin{tabular}[c]{@{}l@{}}Resolution of NLIM ADC\\ (mode)\end{tabular}  & \begin{tabular}[c]{@{}l@{}}5-bit   \\ (PWM )\end{tabular} & \begin{tabular}[c]{@{}l@{}}5-bit   \\  (MCL)\end{tabular} & \begin{tabular}[c]{@{}l@{}}4-bit   \\ (PWM)\end{tabular} & \begin{tabular}[c]{@{}l@{}}4-bit   \\ (MCL)\end{tabular} \\ \hline
Cell number                                                            & 30                                                        & 56                                                        & 14                                                       & 20                                                       \\ \hline
Latency (clock \# ) & 56                                                        & 30                                                        & 20                                                       & 14                                                       \\ \hline
\end{tabular}

 \label{tab:different modes and resolutions}
\end{table}

\begin{table}[t!]
  \centering
  \caption{Comparison of recent ADC in ACIM }
  \resizebox{0.5\textwidth}{!}{ 
\begin{tabular}{|l|l|l|l|l|}
\hline                                                              &\textbf{This   work }                                              & JSSC’23\textsuperscript{\cite{zhang2023macc}} & JSSC’22\textsuperscript{\cite{yu202265}}                                           & JSSC’23\textsuperscript{\cite{lee202328}} \\ \hline
ADC type                                                      & Ramp(IM) & SAR           & Ramp(IM) & Flash           \\ \hline
Technology                                                    & 65 nm                                      & 28 nm               & 65 nm     & 28 nm \\ \hline
Resolution                                                    & \textbf{1-5b}                                                      & 6b           & 1-5b                                                      & 4b              \\ \hline
Linearity                                                     &\textbf{Linear/NL}                                                 & Linear          & Linear                                                    & Linear          \\ \hline
Neuro \#/ADC & \textbf{1}                                                         & 4               & 1                                                         & 16              \\ \hline
MAC array size   &\textbf{160×100}  & 128×128  & 64×128    & 256×80            \\ \hline

 \begin{tabular}[c]{@{}l@{}}Total ADC area   \\ ($A_{ADC},\mu m^2$) \end{tabular}         &\textbf{25650 }   & 6704.2         & 32995       & 7030.8              \\ \hline
 \begin{tabular}[c]{@{}l@{}} One ADC area    \\ (Norm.)* ($\mu m^2$) \end{tabular}     &\textbf{256.5 }   & 141.1    & 257.8   & 947.2  \\ \hline
 \begin{tabular}[c]{@{}l@{}} NL area    \\ (Norm.) ($\mu m^2$) \end{tabular}     &\textbf{26.8 }   & 783.4    & 783.4   & 783.4  \\ \hline
\begin{tabular}[c]{@{}l@{}} One ADC+ NL   \\ area (Norm.) ($\mu m^2$) \end{tabular}    &\textbf{283.3 }   & 924.5    & 1041.2   & 1730.6  \\ \hline

\end{tabular}
}
    \begin{tablenotes}
      {\item*  One ADC area (Norm.)=$A_{ADC}/n_{col}/2^{n-5}$x$AR$. ($n_{col}$ is column \# of MAC array. $n$ is maximum resolution of ADC. $AR$=(65/Technology)\textsuperscript{2}.}
    \end{tablenotes}

\label{tab:ADC_comparison}
\end{table}



\subsection{Error Analysis and Calibration within Hardware }
\begin{figure}[h]
  \centering
  \includegraphics[width= 1\linewidth]{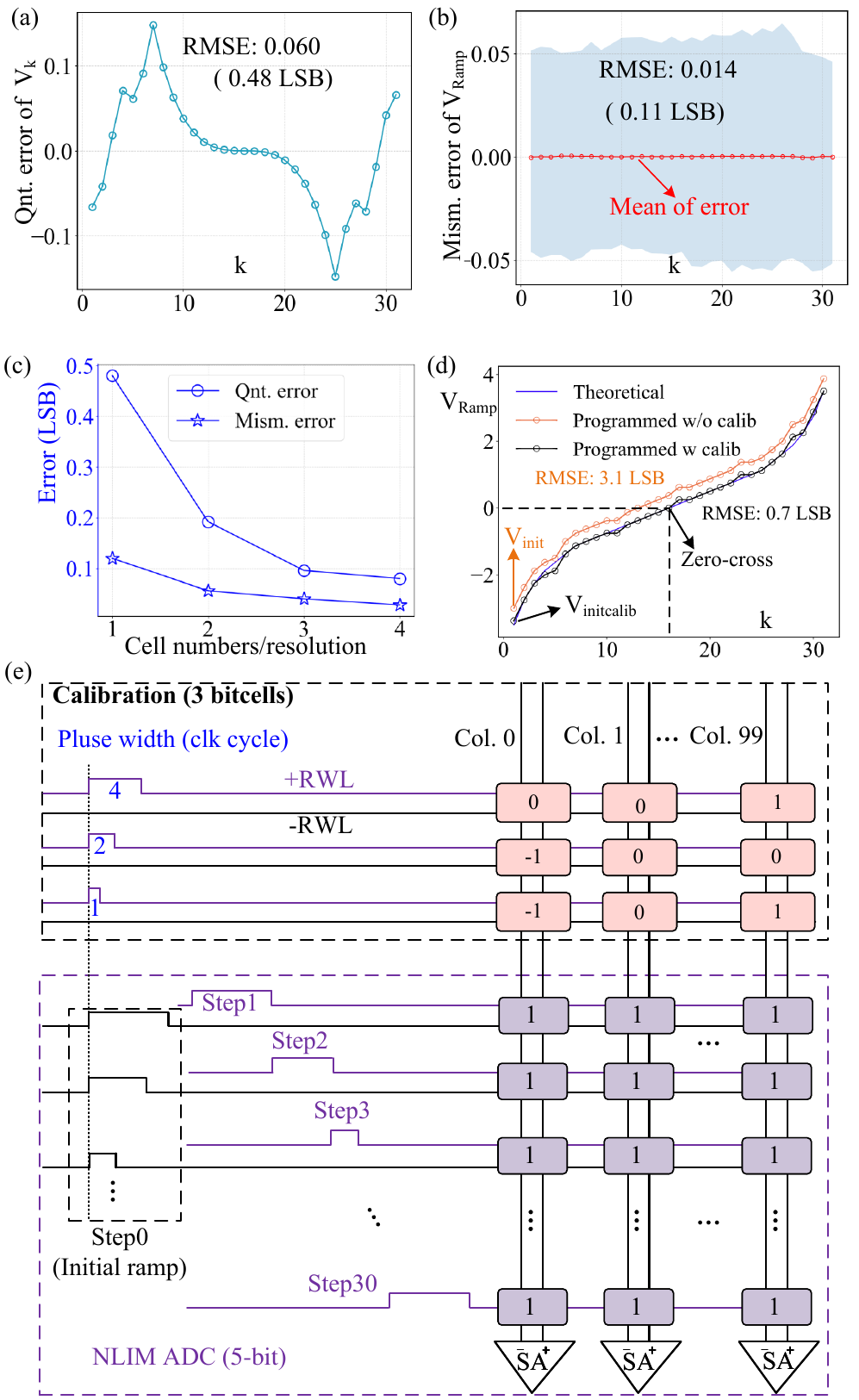}
  \caption{Quantization of $V_k$ and mismatch errors of $V_{Ramp}$ are shown in (a) and (b).  (c) Errors with different resolutions. (d) Zero-cross method for calibration.  (e) Detailed timing diagram of calibration.}
  \label{all_errors}
\end{figure}

The generation of $V_{Ramp}$ requires quantization of $\Delta V_{k}$ as defined in Equation (3), which introduces quantization error. As shown in Fig. \ref{all_errors}(a), the quantization error of $\Delta V_{k}$  leads to an root-mean-square error (RMSE)  of 0.060 (0.48 LSB) generated in $V_{k}$ for the 5-bit NLIM ADC when approximating a sigmoid function. The specific shape of the quantization error curve depends on the function being approximated, but we found the error to be $<1$ LSB for all common NL activations. Our analysis of Equation (3) further reveals that statistical variations among bitcells producing $I_u$ contribute to an additional error named mismatch error. To characterize the mismatch error, we conducted extensive Monte Carlo simulation and developed comprehensive variation model for $I_u$, and then incorporated this model into Equation (3) to generate $V_{Ramp}$ with $I_u$ mismatch. We perform 1,000 simulation runs to statistically quantify the distribution of mismatch error (Fig. \ref{all_errors}(b)). As demonstrated in Fig. \ref{all_errors}(c), both quantization error and mismatch error exhibit an inverse relationship with the number of cells per resolution with quantization error being dominant.

The generated $V_{Ramp}$ deviates from its ideal characteristic due to aforementioned errors. To mitigate this, a zero-crossing calibration technique is employed to adjust the initial voltage $V_{init}$, as illustrated in Fig. \ref{all_errors}(d). The calibration process for the 5-bit ADC (PWM mode) is detailed in Fig. \ref{all_errors}(e), where the initial ramp, calibration, and MAC operations occur simultaneously.

A negative $V_{init}$ is established by simultaneously enabling 15 -RWL pulses in Fig. \ref{all_errors}(e), leveraging the positive weights to generate a negative initial voltage. This enables the ADC to measure both positive and negative MAC values. Subsequently, 30 +RWL pulses are applied sequentially, producing a rising $V_{Ramp}$ (orange line, “Programmed w/o calib" in Fig. \ref{all_errors}(d)). However, the IM ADC in \cite{yue202014} is constrained by its binary input format (only one RWL in every column). To generate $V_{init}$, it must employ a separate array of 15 bitcells. Consequently, this method suffers from a large area overhead, with the $V_{init}$ generation circuit consuming an area comparable to the ADC core itself (IM ADC area/initial ramp area$\approx$50\% \cite{yue202014}), making the cost prohibitively high. 

Calibration utilizes three rows of bitcells with input pulses of 1, 2, and 4 cycles, respectively. Weights in calibration rows are determined experimentally. With MAC and calibration pulses disabled, the ADC output is read at MAC=0. Any deviation from zero indicates a $V_{Ramp}$ offset. For instance, an output of 3 in the first column signifies an upward shift. To realign $V_{Ramp}$ through zero, the corresponding calibration weights are set to (0, -1, -1), shifting $V_{init}$ downward by three resolution steps to $V_{initcalib}$ in Fig. \ref{all_errors}(d). 

This calibration method, applied to determine all weights across the three rows, reduces the RMSE by a factor of 4.4 (from 3.1 LSB to 0.7 LSB in Fig. \ref{all_errors}(d)). It is worth noting that for the 5-bit ADC, we require three additional rows for calibration, leaving 157 rows of bitcells available for MAC computation. However, for a 4-bit ADC, the 10 remaining bitcells can be used for calibration, after 20 are allocated to the ADC module itself as shown in Tab. \ref{tab:different modes and resolutions}. In this case, all 160 rows of bitcells can be used for MAC computation.

\subsection{Robustness to temperature and Weight Calibration}
\label{sec:weight_calib}
\begin{figure*}[htbp]
  \centering
  \includegraphics[width= 1\linewidth]{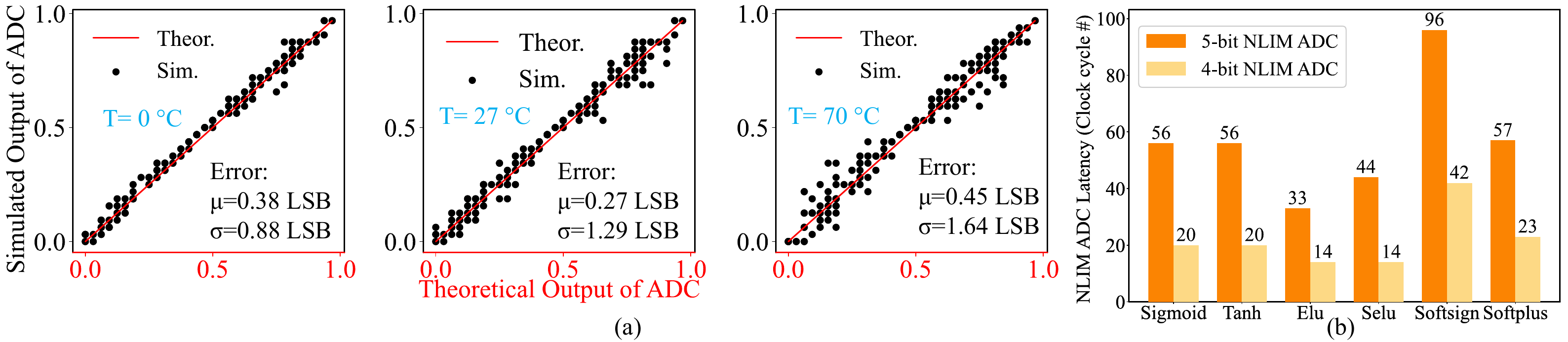}
  \caption{(a) Robustness of our 5-bit NLIM ADC implementing sigmoid function under different temperatures (2-bit weight). (b) Total latency of NLIM ADC for different NL activations.}
  \label{temperature_and6latency_sim}
\end{figure*}
Fig. \ref{temperature_and6latency_sim}(a) shows the distribution of the 5-bit NLIM ADC output implementing a sigmoid function compared to theoretical output values  using data from GSCD with 2-bit weight quantization for LSTM weights for $0^\circ$C, $27^\circ$C and $70^\circ$C. It can be observed that the error remains consistently small in Fig. \ref{temperature_and6latency_sim}(a)) demonstrating the robustness of the NLIM ADC architecture against temperature fluctuations due to replica bias (the bitcells in NLIM ADC section used to generate the NLIM ADC references are identical to those in MAC section). When this maximum error distribution (N(0.45 LSB, 1.64 LSB) at $70^\circ$C) is added to the LSTM network, the simulated accuracy drops by only 1.1\% (from 91.2\% to 90.1\%). 


The $\Delta VDD_{CORE}$ in Section \ref{sec:Implementation of Multi-bit Weight} exhibits positive temperature dependence due to $U_T$. To maintain bit weighting ratio $n_{BWR} = I_{MSB}/I_{LSB} \approx 2$ for our 3-bit LSTM network, we implement a temperature-proportional current ($I_{PTAT}$)-based voltage generator \cite{mandal2006fast}, \cite{delbruck2005bias}. Using $I_{PTAT}$ to create $V_{MSB} = V_{LSB} + I_{PTAT}R$ (with fixed $V_{LSB}=0.42V$ and $R$ ). This design maintains a nearly constant $n_{BWR}=e^{\kappa R \frac{I_{PTAT}}{U_T}}$ across temperature variations, since both the $I_{PTAT}$ and $U_T$ increase proportionally with temperature. SPICE simulation shows $n_{BWR}$ remains within 1.99-2.02 across 0-70$^\circ$C. The worst case  $n_{BWR}$ results in merely 0.15\% inference accuracy degradation for 3-bit weight LSTM network.

In case a more precise ratio is necessary, we can do a simple calibration routine to make small adjustments to the $\Delta VDD_{CORE}$. Calibration can be performed by making the resistance, $R$, in the reference generator programmable (can be a combination of fixed and digitally tunable $R$). For calibration, the same set of inputs are applied to the array twice, while keeping the weights in the MSB and LSB array as ($W_{MSB},W_{LSB}$)=(0, $W_{cal}$) and ($W_{cal}$,0) respectively, where $W_{cal}$ denotes a set of random weights used for calibration. The ADC is configured as a linear one and the measured outputs for the two cases can be used to extract an accurate estimate of the actual $n_{BWR}$ on-chip. Next, binary search can be used to find the ideal resistor value to tune this ratio close to the desired one (=2 for 3, 4-bit weight, =4 for 5-bit weight).

 \subsection{Latency analysis}
 Our NLIM ADC can approximate different NL activations. The latency of 5-bit and 4-bit NLIM ADC with PWM mode for different NL activations is shown in Fig. \ref{temperature_and6latency_sim}(b), obtained according to the equation $\sum_{k=1}^{30} Qnt.(\Delta V_{k})$.




\section{Measurement and Simulation results}
\label{sec:Measurement and simulation results}
A $0.92\times 0.92$ mm$^2$ chip containing the macro was fabricated in a $65$ nm CMOS process occupying a core area of 0.23 mm$^2$.

\subsection{Programmed In-memory Nonlinear ADC Results}
\label{sec:Programmed In-memory Nonlinear ADC Results}
\begin{figure*}[t]
  \centering
  \includegraphics[width=1\linewidth]{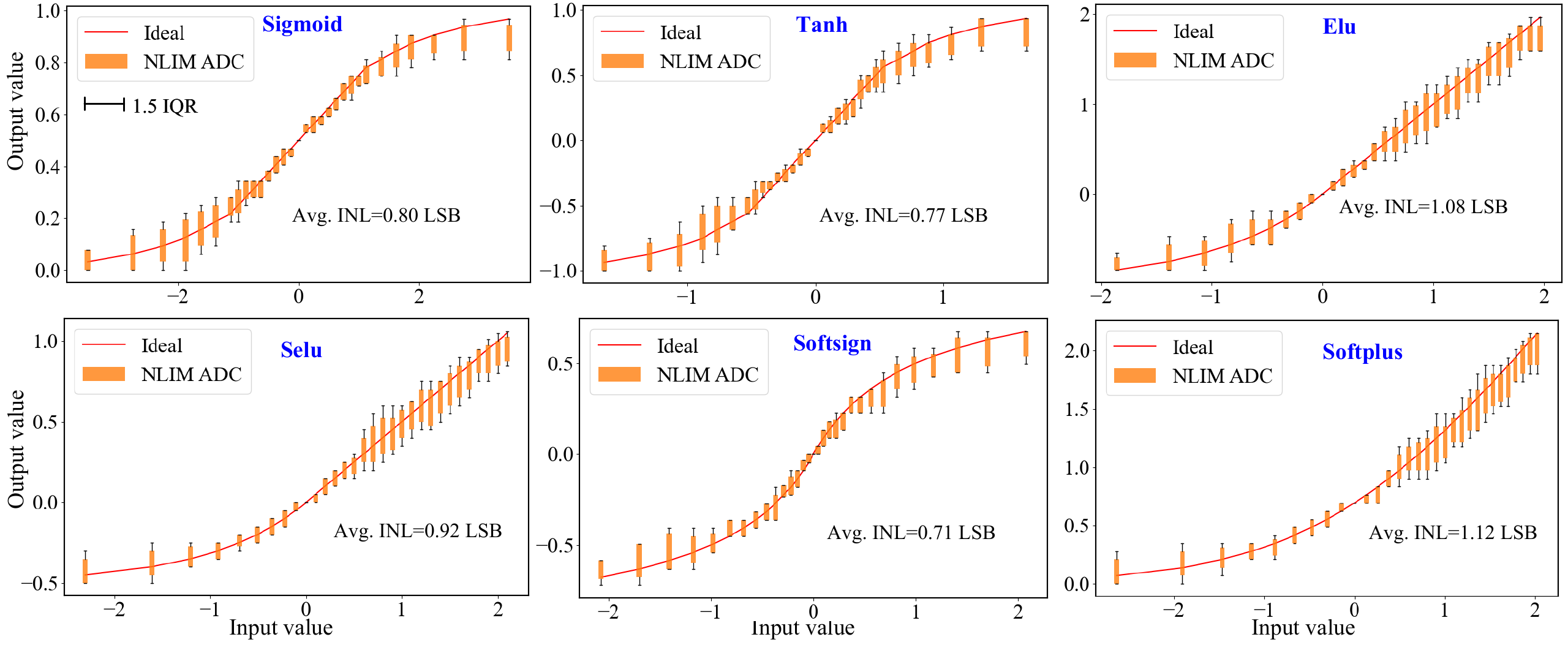}
  \caption{Experimental results of approximating six  NL activations at 5-bit resolution. ($\text{Avg. INL} =\sum_{col=1}^{100}[(\sum_{i=1}^{31}\mid  Ideal_{col,i}- NLIM ADC_{col,i}\mid) / 31)]/100$)  }
  \label{three_NL_measrured_result}
\end{figure*}

We experimentally programmed six common neural network NL activations (Sigmoid, Tanh, Elu, Selu, Softsign and Softplus) in our chip (100 columns) using 5-bit NLIM ADC with calibration, as shown in Fig. \ref{three_NL_measrured_result}. The average integral non-linearity (INL) for approximating these six NL activations are 0.80 LSB, 0.77 LSB, 1.08 LSB, 0.92 LSB, 0.71 LSB and 1.12 LSB, respectively. 

\subsection{LSTM for KWS Task}
\label{sec:KWS Accuracy Results}
\begin{figure*}[t]
  \centering
  \includegraphics[width=\linewidth]{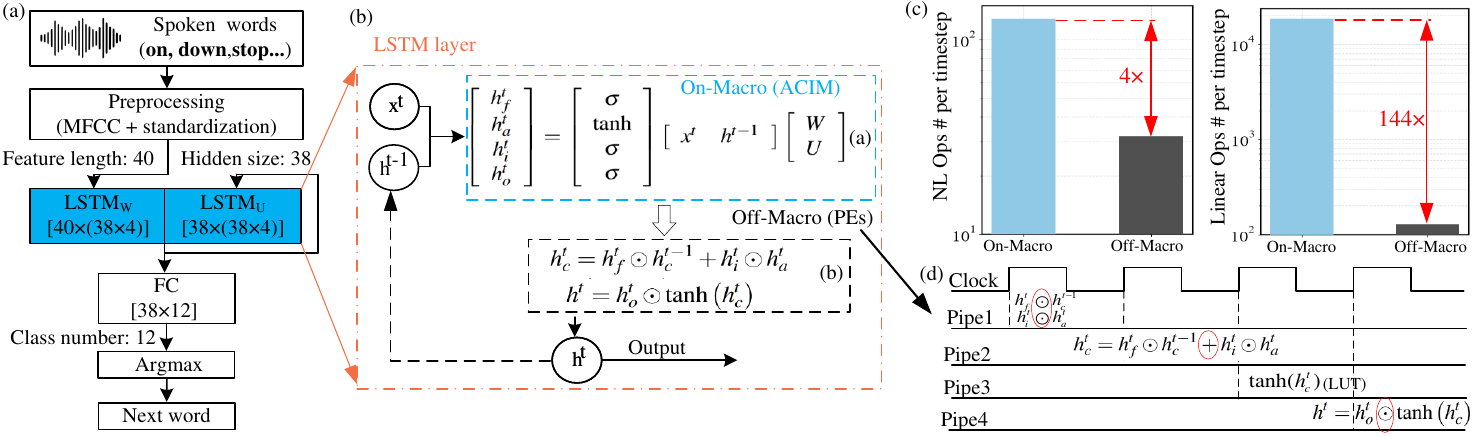}
  \caption{(a) Architecture of LSTM network on-chip inference. (b) Mapping of the LSTM network onto the
Macro of chip. (c) Operations breakdown in the LSTM layer of one timestep. 80\% nonlinear operations and 99\% linear operations can be done on-macro.  (d) Pipe-lined timing diagram for PEs (off-macro).}
  \label{LSTM for KWS task}
\end{figure*}

The GSCD dataset\cite{warden2018speech} is used to train and test for the 12-class KWS task. The KWS  training network includes  Mel-frequency
cepstral coefficient (MFCC) for feature extraction, standardization, LSTM layer  and fully connected (FC) layer. Quantization and hardware nonidealities aware training (HWAT) \cite{hou2019normalization},\cite{mao2022experimentally},\cite{ joshi2020accurate} is employed to enhance the resilience of weights against quantization error and hardware nonidealities so that the network can be tested on-chip with less loss of accuracy. Our LSTM network can achieve better performance, even when employing low-bit quantization (2 or 3 bits) for weights using the quantization method as introduced in \cite{hou2019normalization}. After 2-bit and 3-bit quantization, 60\% and 67\% of the weights in the macro are zero, respectively. These zero-valued weights, as illustrated in Fig. 2(d) (middle), create no discharge paths within the bitcell, thereby reducing the energy consumed by RBL discharge.

On-chip inference depicted in Fig. \ref{LSTM for KWS task}(a) is performed after training.  After feature extraction and standardization (both procedures are implemented off-chip), These extracted features, along with the previous 38-dimensional output vector $h^{t-1}$ from the LSTM, are then sent to the Macro in the chip.  Fig. \ref{LSTM for KWS task}(b) illustrates the division of operations within the LSTM layer, specifying which components are executed within the Macro of our chip and which are processed externally. The parameters (W and U) and NL activations (sigmoid ($\sigma$) and tanh) in Equation (a) in Fig. \ref{LSTM for KWS task}(b) are programmed in the 6T-SRAM array of ACIM while $h^{t-1}$ and $x^{t}$ are converted into PWM pulses in Macro as the input of ACIM.  While the Macro does not support the full set of LSTM operations for this model, it efficiently handles 80\% of NL operations and 99\% of linear (MAC) operations in the LSTM layer as shown in Fig. \ref{LSTM for KWS task}(c). For the computations in Equation (b) in Fig. \ref{LSTM for KWS task}(b), we implement a pipelined digital circuit composed of processing elements (PEs), achieving a latency of 4 clock cycles per data set as shown in Fig. \ref{LSTM for KWS task}(d). The system employs 19 parallel PEs, with each PE processing 2 out of the total 38 dimensions of the hidden state, resulting in an overall latency of 5 cycles.  Using TSMC 65 nm standard digital cells for synthesis, we estimate a total power consumption of this part as 32 $\mu W$ at 100 MHz, 1 V.

\begin{figure*}[htbp]
  \centering
  \includegraphics[width=1\linewidth]{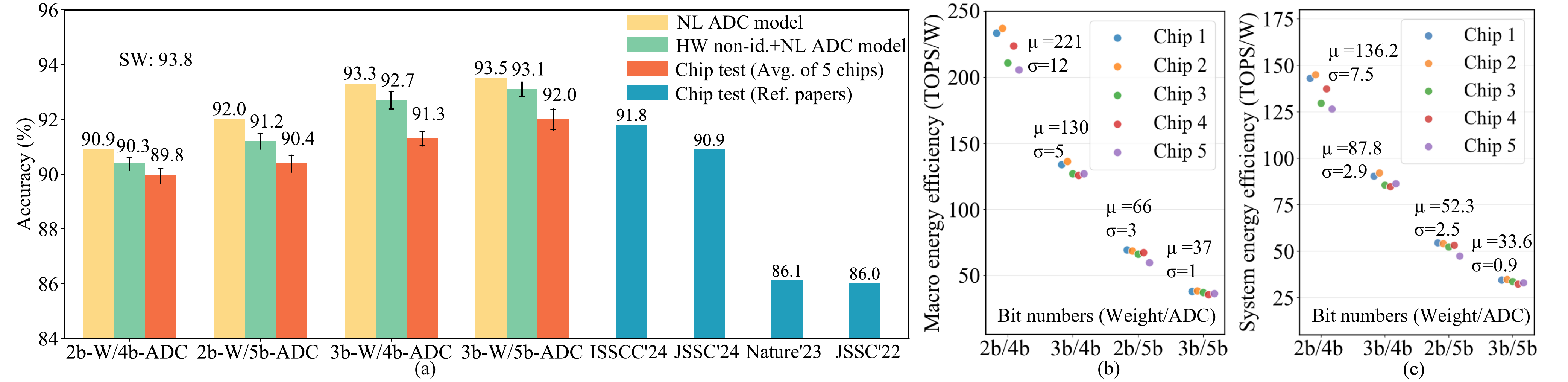}
  \caption{(a) Comparison of the our inference accuracy results with the state-of-the-art results of 12 classes (ISSCC'24\cite{tan202417}, JSSC'23\cite{tan20230}, Nature'23\cite{ambrogio2023analog}, ISSCC'22\cite{kim202223muw}). (b) Macro energy efficiency (including nonlinear operations) of five chips under different resolutions @ 100 MHz 0.45V, 0.8V, 1V for 2-bit weight ($VDD_{core}=$0.45V, $\Delta VDD_{core}=$ 0.03 V, $V_{RWL}=$0.8V, $VDD=$1V for 3-bit weight). (c) System energy efficiency (including LSTM and FC layers).}
  \label{Test_accuracy}
\end{figure*}
Inference accuracy results in software for 12-class KWS is shown in Fig. \ref{Test_accuracy}(a). After adding the NLIM ADC model to replace the NL functions in the LSTM layer, inference accuracies for the four configurations—employing 2-bit and 3-bit weights with 4-bit and 5-bit NLIM ADC—are  90.9\%, 92.0\%, 93.3\% and 93.5\%, respectively, which compare favorably with a floating-point baseline of 93.8\%. 
To improve the robustness of the neural network against hardware non-ideality when mapping weights to the chip, we incorporated a simple hardware nonideality model N(0,0.05) (obtained from actual chip measurements similar to Fig. \ref{temperature_and6latency_sim}(a)) to the output of NLIM ADC during the training process. Consequently, the software-based inference accuracies change to  90.3\%, 91.2\%, 92.7\% and 93.1\%, respectively. The robustness of the classification was verified through 10 independent experimental runs, resulting in a small standard deviation (Fig. \ref{Test_accuracy}(a)).  Next, we experimentally evaluated the on-chip inference accuracy variations of five chips across four different configurations, observing mean accuracy drop within $\approx$1\% margin of the hardware nonidealities calibrated NLIM ADC software model and low variability across five chips. As shown in Fig. \ref{Test_accuracy}(a), the averaged on-chip inference accuracy values (with error bars of five chips) under these four configurations are 89.8\%, 90.4\%, 91.3\% and 92.0\%, respectively. Our averaged on-chip inference accuracy is 92.0\% for 5-bit ADC and 3-bit weight, demonstrating competitive performance with the state-of-the-art result of 91.8\% \cite{tan202417}. 

The maximum accuracy drop in Fig. \ref{Test_accuracy}(a) between the model (accounting for hardware nonidealities and  ADC model) and the actual hardware is 1.4\%, decreasing from 92.7\% (green bar) to 91.3\% (orange bar) at 3-bit weight and 4-bit ADC. The primary reason for this accuracy drop is the simple non-ideality model we have used during training which added noise from a fixed error distribution obtained after testing a fixed set of input data. In reality, the approximation error of the NL activation varies with different input data (note different variability in the different points of Fig. \ref{three_NL_measrured_result}). In future, we will explore adding more detailed non-ideality model of the ADC as well as mismatch induced variations of $n_{BWR}$ to have a better model of hardware nonidealities aware training.
\subsection{Hardware Performance}
\label{sec: Hardware Performance}

\begin{figure}[t!]
  \centering
  \includegraphics[width=1\linewidth]{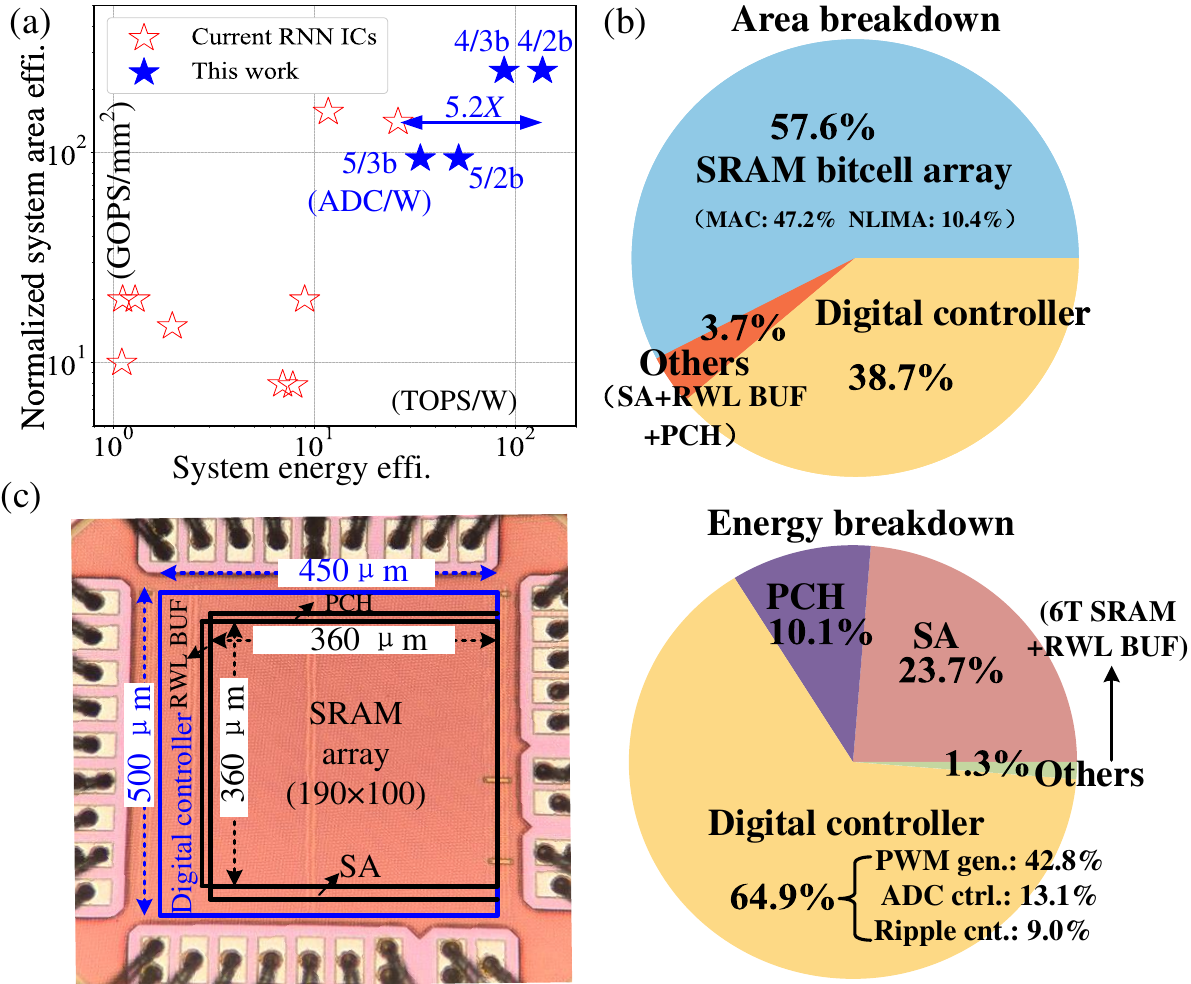}
  \caption{ (a) Comparison of our work with other RNN ICs. (b) Chip area and measured energy breakdowns (c) Die photograph of total active area 0.23 mm\textsuperscript{2}.}
  \label{Hardware Performance}
\end{figure}


\begin{table*}[!t]
\begin{center}
\resizebox{1\textwidth}{!}{ 
\begin{threeparttable}
  \caption{Performance Comparison of RNN accelerators with state-of-the-arts }
\begin{tabular}{|l|l|l|l|l|l|l|l|l|}
\hline
Metric                                                                       & VLSI’19\textsuperscript{\cite{guo20195}} & JSSC’20\textsuperscript{\cite{kadetotad20208}} & VLSI’21\textsuperscript{\cite{guo20216}} & JSSC’21\textsuperscript{\cite{dbouk20200}} & Nat. Elec.’23\textsuperscript{\cite{le202364}} & Nature’23\textsuperscript{\cite{ambrogio2023analog}} & JSSC’23\textsuperscript{\cite{ tambe202216}}  & This work   \\ \hline
CMOS technology                                                              & 65 nm           & 65 nm           & 65 nm      & 65 nm      & 14 nm                & 14 nm             & 16 nm                       & 65 nm       \\ \hline
Supply   voltage(V)                                                          & 0.9         & 0.68        & 0.75   & 0.7, 1         & 0.85-0.95            & 0.8-1.8           & 0.8                           & 0.42/0.45, 0.8, 1      \\ \hline
Memory technology                                                            & SRAM            & SRAM            & SRAM       & SRAM      & PCM                  & PCM               & SRAM                        & SRAM        \\ \hline
Bitcell                         & 6T            & --            & 6T       & 6T      & 8T4R                  & 6T4R               & --                        & Dual 9T        \\ \hline
Frequency (MHz)                                                              & 75              & 80              & 160       & 500      & 1000                 & 1000              & 573                           & 100         \\ \hline
In./w./out. precision                                                        & 1/1/3           & 13/6/13         & --/2/2    & 8/8/8      & 8/Analog/8             & 8/Analog/8          & 8/8/8                      & 4-5/2-3/4-5 \\ \hline
CIM inclluding NL                                                                          & N               & N               & N          & N      & N                    & N                 & N                             & Y           \\ \hline
Algorithm                                                                   & RNN             & LSTM            & LSTM       & RNN         & LSTM                 & LSTM              & RNN                        & LSTM        \\ \hline
Dataset                                                                   & GSCD             & TIMIT            & GSCD      & GSCD       & PTB                 & GSCD              & LibriSpeech                           & GSCD        \\ \hline
Acc.(\%)/class num.                                                      & 90.2/10           & --/--            & --/--      & 90.38/7        & --/--                 & 86.1/12              & --/--                          & \begin{tabular}[c]{@{}l@{}}89.8 (4/2/4b)/12\\    92.0 (5/3/5b)/12 \end{tabular}       \\ \hline
Power (mW)                         & 52.51            & 1.85            & 46.90       & --      & 3450                  & 3317               & 214                        & \begin{tabular}[c]{@{}l@{}}0.41 (4/2/4b)\\     0.63 (5/3/5b) \end{tabular}        \\ \hline

\begin{tabular}[c]{@{}l@{}}Sys.   energy\\  effi. (TOPS/W)\end{tabular} & 11.7            & 8.93\textsuperscript{(b)}            & 26.03       & 0.91      & 1.96                 & 6.94              & 7.8                       & \begin{tabular}[c]{@{}l@{}}136.2 (4/2/4b)\\     33.6 (5/3/5b) \end{tabular}      \\ \hline
\begin{tabular}[c]{@{}l@{}} Norm. sys. energy \\  effi. (TOPS/W)\textsuperscript{(c)}\end{tabular} & 11.7            & 696.5\textsuperscript{(b)}            & 104.12       & 58.24      & 125.44\textsuperscript{(d)}                 & 444.16\textsuperscript{(d)}              & 499.2                       &     1089.6\textsuperscript{(e)}        \\ \hline

\begin{tabular}[c]{@{}l@{}}Sys.   area effi. \\ (GOPS/mm\textsuperscript{2})\end{tabular} & 156             & 20    & 228  & --            & 320                  & 170               & 130                          & \begin{tabular}[c]{@{}l@{}}245 (4/2/4b)\\    93.7 (5/3/5b) \end{tabular}         \\ \hline
\begin{tabular}[c]{@{}l@{}}Norm. sys. area  \\ effi. (GOPS/mm\textsuperscript{2})\textsuperscript{(a)}\end{tabular} &156      &20         & 228    & --             & 14.9                     &7.9                   &7.8                                & \begin{tabular}[c]{@{}l@{}}245 (4/2/4b)\\    93.7 (5/3/5b) \end{tabular}            \\ \hline
\end{tabular}

\begin{tablenotes}
    \item(a) Normalized  system area efficiency=  system area efficiency × (process/65 nm)\textsuperscript{2};  (b) FC layer is not included; (c) All energy efficiencies are normalized to 1-bit input and 1-bit weight, according to this formula \cite{song20234}: Normalized EE=EE × input precision × weight precision; (d) The analog weights are normalized using 8 bits. (e) The data is normalized from 4/2/4b resolutions.
\end{tablenotes}
  \end{threeparttable}
}
\end{center}
\label{tab:RNN_comparison}
\end{table*}

Fig. \ref{Test_accuracy}(b) presents the measured macro-level EE (including NL operations) of our design for the aforementioned four configurations, ranging from 37 TOPS/W to 221 TOPS/W. The measured fluctuation of macro EE remains within 10\%. Following the evaluation methodology (pipe-lined architecture) described in Section \ref{sec:KWS Accuracy Results}, we estimate the latency, power consumption, and area of the FC layer. The resulting system-level EE and area efficiency for inference are [33.6, 52.3, 87.8, 136.2] TOPS/W as shown in Fig. \ref{Test_accuracy}(c) and [93.7, 93.7, 245, 245] GOPS/mm\textsuperscript{2} under above four configurations, respectively. As shown in Fig. \ref{Hardware Performance}(a), when comparing our LSTM accelerator with current RNN accelerators \cite{conti2018chipmunk,yin20171,shin201714,tambe202216,ambrogio2023analog,le202364,dbouk20200,guo20216,kadetotad20208,guo20195}, our design achieves 5.2× higher  system-level EE and 1.6× better normalized (65 nm) area efficiency  than SOTA solutions\cite{guo20216,guo20195} when operating at 2/4 bits (bit numbers: weight/NLIM ADC). 

Fig. \ref{Hardware Performance}(b) displays the detailed area and energy breakdowns at supply voltages of $VDD_{core}=$0.45V, $\Delta VDD_{core}=$ 0.03 V, $V_{RWL}=$0.8V, $VDD=$1V and 100 MHz clock frequency at 3/5 bits. The NLIM ADC (bitcell+SA) module occupies only 11\% of the total area. The measured energy breakdown reveals that the digital controller dominates power consumption (64.9\%), while the NLIM ADC module accounts for approximately 36.8\% of total energy. Fig. \ref{Hardware Performance}(c) presents the die micrograph with total active area 0.23 mm\textsuperscript{2}. Tab. VI provides a comprehensive performance comparison between our LSTM  accelerator and SOTA RNN accelerators, highlighting improvements in configuration flexibility (4-5/2-3/4-5b, input/weight/output), accuracy of KWS,  energy/area efficiency, and CIM including NL. For a fair comparison, all energy efficiency values are normalized to the configuration of 1-bit weight and 1-bit input, following the formula in \cite{song20234}. Our normalized system energy efficiency (1089.6 TOPS/W, including the FC layer) is 2.2× higher than the SOTA result (499.2 TOPS/W reported in \cite{ tambe202216}). While we used a relatively older process of 65 nm CMOS, we expect even better performance in terms of leakage power, mismatch and bitcell area when moving to advanced nodes such as 22 nm FDSOI due to the intrinsically better characteristics of the transistor \cite{fdsoi-ref}.

\section{Conclusion}
\label{sec:Discussion and Conclusion}

In this work, we propose the ACIM macro supporting reconfigurable NLIM ADC to approximate NL activations directly in the analog domain. Experimental results show the 5-bit NLIM ADC for approximating NL activations in LSTM cells, achieving $<1$ LSB average error with a 3.3×/3.7× normalized area reduction over previous SAR ADC and linear IM ADC including NL units. Simulations further demonstrate its temperature robustness enabled by replica bias. Our LSTM accelerator achieves 92.0\% on-chip inference accuracy for 12-class KWS task while demonstrating 2.2× higher system-level normalized EE and 1.6× better normalized area efficiency compared to SOTA designs. A dual 9T bitcell is proposed to support signed inputs and ternary weights, combined with our RUDC technique that increases DR of $V_{RBL}$ by 2.8× versus prior implementation.  Dual-supply 6T-SRAM array is employed for multi-bit weight implementation, reducing both bitcell count (7.8×) and latency (4×) for 5-bit weight operations.

\section{Appendix}

\subsection{Model training}
The neural network training consists of three parts: data preprocessing and LSTM layer and  fully connected (FC) layer.

During the data preprocessing phase, the GSCD \cite{warden2018speech} dataset serves as the foundation for model development. This comprehensive dataset comprises 65,000 one-second audio clips, collected from numerous individuals through public contributions. The original audios are consolidated into 12 primary classifications. Each one-second audio segment contains 16,000 discrete sampling points. Mel-Frequency Cepstral Coefficients (MFCC) processing is employed to extract essential vocal characteristics, with 49 analysis windows segmenting each audio recording to derive 40 distinctive features per window.

The LSTM training phase incorporates a customized LSTM layer (78×152=11856 weight parameters) as its central computational element.  The network configuration processes input vectors of length 40 while maintaining a hidden state dimension of 38. With a sequence length of 49 and batch size of 1024, each training iteration involves 49 sequential computations within the LSTM architecture. Our LSTM network can achieve better performance, even when employing low-bit quantization (2 or 3 bits) for weights because of  quantization-aware training (QAT) \cite{hou2019normalization}. The following Equation (4) is an example for 2-bit weight quantization. 

\begin{equation}
W_k =
\begin{cases}
+1, & \text{if } W_k > 0.7m\\[2pt]
0,  & \text{if } -0.7m \le |W_k| \le 0.7m\\[2pt]
-1, & \text{if } W_k < -0.7m.
\end{cases}
\end{equation}
where the threshold $m$ is the average of the absolute values of all the weights.

 Algorithm 1 formalizes the  hardware nonidealities aware training (HWAT)\cite{mao2022experimentally},\cite{ joshi2020accurate}  process, which employs hardware-based nonidealities sampled from a normal distribution. HWAT is utilized to enhance the resilience of weights against  hardware nonidealities so that the network can be tested on-chip with less loss of accuracy.
\begin{algorithm}[!t]
\caption{Hardware nonidealities aware training (HWAT)}\label{alg:alg1}
\begin{algorithmic}
\STATE 
\STATE Input: $x^t$, $h^{t-1}$, $W_t$ and $Nonidealities$
\STATE Output: $W_{t+1}$
\STATE \textbf{loop}
\STATE \hspace{0.4cm} Feed forward propagation: 
\STATE \hspace{0.5cm}$ Y_\mathbf{t} \gets  Q([x^t,h^{t-1}] \cdot W_t) $
\STATE \hspace{0.5cm}$ Z_\mathbf{t} \gets  f(Y_t) +Nonidealities $
\STATE \hspace{0.4cm} Back-propagation: 
\STATE \hspace{0.5cm}$ g_\mathbf{t} \gets$  calculate gradient based on $W_t$ and $f(Y_\mathbf{t})$
\STATE \hspace{0.5cm}$ W_\mathbf{t+1} \gets W_\mathbf{t}+\eta \times g_\mathbf{t} $
\STATE \textbf{end loop}
\end{algorithmic}
\end{algorithm}

A FC layer subsequently processes the 38-dimensional feature vectors generated by the LSTM to perform the final 12-class classification. The model is optimized using the Adam optimizer and a cross-entropy loss function. 



%


\bibliography{IEEEexample}
\bibliographystyle{IEEEtran}

\end{document}